# Global Supply Chain Reallocation and Shift under Triple Crises: A U.S.-China Perspective

Wei Luo[1][a]*, Siyuan Kang[1a], Qian Di[b]

[a] GeospatialX Lab, Geography Department, National University of Singapore, Singapore, 117568

[b] Vanke School of Public Health, Tsinghua University, Beijing

*Corresponding Authors:

Wei Luo | Geography Department, National University of Singapore, Singapore, 117568 | Office Phone: (65) 6516-3851 | Email: geowl@nus.edu.sg

**Abstract:**

US-China trade tensions, the COVID-19 pandemic, and Russia-Ukraine conflict have disrupted global supply chains for reallocation. Existing studies raise cautions that US-China trade tensions may not lead to the reduction of US dependence on supply chains linked to China. What are the potential driving forces behind this unmet reallocation away from China in the context of those overlapping geopolitical and public health disruptions. This study investigates how the above three events jointly reconfigured bilateral trade and global value chain (GVC) participation and positioning among U.S., China, and top trading partners between 2016 and 2023. Leveraging monthly bilateral trade data across those countries and all economic sectors, together with multi-regional input–output tables for GVC decomposition, we combined a multi-period event-study framework with a structural analysis to evaluate both trade flow disruptions and shifts in participation and functional positioning within GVC. We found that China's exports remained robust, expanding across global markets while keeping sustained rise in GVC participation and more embedded in upstream GVC segments through increased intermediate shipments to Asia and Europe. Meanwhile, U.S. imports increasingly shifted toward "China+1" partners, such as Associations of Southeast Asian Nations —whose trade structures remain closely relied on the Chinese upstream supply chains. The strengthening and embedded triangular trading relationships provide more evidence on how the global reallocation and GVC evolve, with a focus on the US and China over successive global shocks. Through this study, we propose a novel supply chain resilience framework that consists of the interaction of three key dimensions: the level of GVC participation, the functional position within the value chain, and the country's capacity to re-couple in the post-shock landscape, which depends on market diversification, economic complexity, and institutional capability. These findings carry significant implications for trade policy and industrial strategy in an era of geopolitical and geoeconomic fragmentation.

## Introduction

Global supply chains have undergone significant transformations over the past few decades, driven by advances in technology, evolving economic policies, and shifting geopolitical landscapes [1–3]. This transformation has fostered an intricate web of trade relationships – often referred to as global value chains (GVCs) – where interdependent economies amplify the ripple effects of policy and market changes [4,5]. This complex interdependence is particularly evident in the trade dynamics between major economic powers such as China and the United States [6–8]. Recent events, such as the U.S.-China trade tensions, the COVID-19 pandemic, and the Russia-Ukraine conflict, have underscored the fragility and adaptability of this interconnected system, collectively introducing new variables into GVCs [9–12].

The U.S.-China trade tensions, beginning in 2018, initiated significant shifts in global supply chains as industries sought to mitigate risks from escalating tariffs [10–13]. Companies diversified production networks and sourcing strategies to reduce dependence on single markets, prompting a rethinking of supply chain resilience [14,15]. The COVID-19 pandemic, with its widespread disruptions to production, logistics, and demand, further underscored the vulnerabilities in global trade systems [16–18]. The crisis heightened awareness of the need for supply chain adaptability, especially for critical goods such as medical supplies and pharmaceuticals [19,20]. More recently, the Russia-Ukraine conflict has intensified concerns over energy security and access to essential commodities like natural gas and grains, driving nations to reconsider strategic dependencies [21,22]. Together, these events have catalyzed fundamental reallocations of global supply chains and the balance between economic efficiency and national security.

The reconfiguration of GVCs is fused at a deeper level of a broader reevaluation of globalization's benefits and risks, particularly within the United States [23,24]. These events have exposed vulnerabilities in GVCs, prompting a strategic shift in U.S. policy and corporate practices aimed at reducing dependence on Chinese supply chains, particularly for goods deemed critical to national or economic security. Those measures reflect a growing emphasis on "reshoring", "friendshoring" and "nearshoring" [25]. Research indicates that the US-China trade tensions caused the U.S. to import less from China directly with Vietnam as friendshoring alternative and Mexico as nearshoring one [26]. However, it still raises some cautions whether those policies will finally reduce the dependency of U.S. on Chinese supply chains because the US could remain indirectly connected with China through importing from Vietnam and Mexico [27]. The evidence of reallocation away from China is not consistent [28]. For example, the reliance of the global supply chains on the exports of China and Asia-Pacific increased significantly during the COVID-19 because of their effective disease control at an early stage[29,30]. Based on the presented premises and the product and market diversification theory supported by empirical work[31,32], we expect that both countries will inevitably reduce external vulnerability via expanding into new economic sectors and markets to improve economic resilience.

But U.S. policy aside, China's manufacturing and export capabilities are crucial in global supply chains [33]. As a global manufacturing hub, China supplies a significant amount of intermediate and final products, supporting production activities in many countries [34]. Recent studies explored China's integration with Southeast Asian supply chains, demonstrating China's influence in enhancing regional supply chain efficiency. For instance, trade cooperation between China and ASEAN countries in electronics, automotive parts, and textiles has become increasingly close, promoting regional economic growth and enhancing supply chain resilience [35]. Other analyses have examined technological and supply chain cooperation between China and other Asian countries, highlighting the contribution of such cooperation to regional economic resilience [36]. For example, cooperation between China, South Korea, and Japan in semiconductor and new energy sectors has driven technological innovation and industrial upgrading [37].

Recent research underscores the significance of both GVC participation levels—captured through metrics such as foreign value added and domestic value embedded in partners' exports—and functional positions in shaping countries' vulnerability to external disruptions[38,39]. Economies positioned upstream, primarily supplying intermediate goods, tend to exhibit greater resilience to demand-side contractions due to their foundational role in production, though they remain vulnerable to supply-side disruptions such as raw material shortages or input bottlenecks[40]. In contrast, countries integrated at downstream stages have benefited from trade diversion amid recent geopolitical tensions but continue to occupy relatively limited roles in assembly-based segments[41]. Beyond functional positioning, participation depth also matters, diversified industrial structures are often more embedded and stable within GVCs. For example, China's growing role as both a major source of intermediate goods and a consumer market has translated into a steady rise in GVC participation, particularly in manufacturing and high-tech sectors. Evidence from recent global disruptions shows that China's backward GVC engagement remained relatively stable compared to forward linkages—suggesting that its vast internal supply ecosystem buffers against external decoupling pressures[42]. Combining the role of GVC's exposure to shocks and observed facts, we expect to observe the trend of increasing GVC participation and upstream for Chinese manufacturing and exports in this era of radical uncertainty.

While prior research has highlighted the macro-level reallocation of global supply chains in response to exogenous shocks such as the U.S.–China tariffs, empirical evidence has largely remained confined to changes in aggregate trade volumes or re-export flows[26]. Furthermore, most focus on the impact of single event (i.e., US-China trade tensions) on several major trading countries (i.e., US, China, and third-party trading partners) without considering the global trading patterns in the context of the triple crises. It is key to providing empirical studies to address the above research gaps for reassessing those trading policies and evaluating the resilience of global supply chains when facing destructive events. It may reveal the potential driving forces behind the unmet original goal of allocation away from China in U.S. policy and corporate practices. It can also provide insights into the potential consequences of the Trump administration's imposition of steep protective tariffs in 2025 and suggest supply chain strategies for the rest of the world.

To address the above questions, this study provides a comprehensive and integrated analysis of trade dynamics among China, the United States, and the top 20 third-party trading partners over the period from 2016 to 2023—a timeframe shaped by the convergence of three major global disruptions leading to the global reallocations of GVCs. This research moves beyond conventional trade flow metrics by jointly examining import and export activities of both countries across all economic sectors. Using an event-study framework, we traced how each shock affected bilateral trade flows across sectors and destinations. Crucially, we extend this analysis from a GVC perspective—investigating not only trade volumes but also the evolving functional roles that China, the United States, and key third-party trading partners occupy within international production systems. By applying decomposition-based indicators of GVC participation and position, the study captures how these roles—whether upstream suppliers of intermediates or downstream assemblers of final goods—have shifted under external pressures.

Building on these foundational insights, our study extends the analytical frontier by integrating multi-scalar event study methods with value-added decomposition to reveal how bilateral and third-party trade dynamics evolve within the architecture of global value chains. Beyond documenting shifts in trade direction, we uncover the functional reconfiguration of economies within GVCs—capturing not only where trade is diverted but how countries adapt in terms of production stages and value-added roles. These findings provide new empirical grounding for conceptual frameworks of supply chain resilience, showing that resilience is not merely a return to pre-shock configurations, but often entails strategic repositioning, intermediation, and long-term structural adjustment. In doing so, our results offer a more granular and functional interpretation of resilience amid global uncertainty.

# Results

## General analysis：Macroeconomic shifts in trade patterns under global disruptions

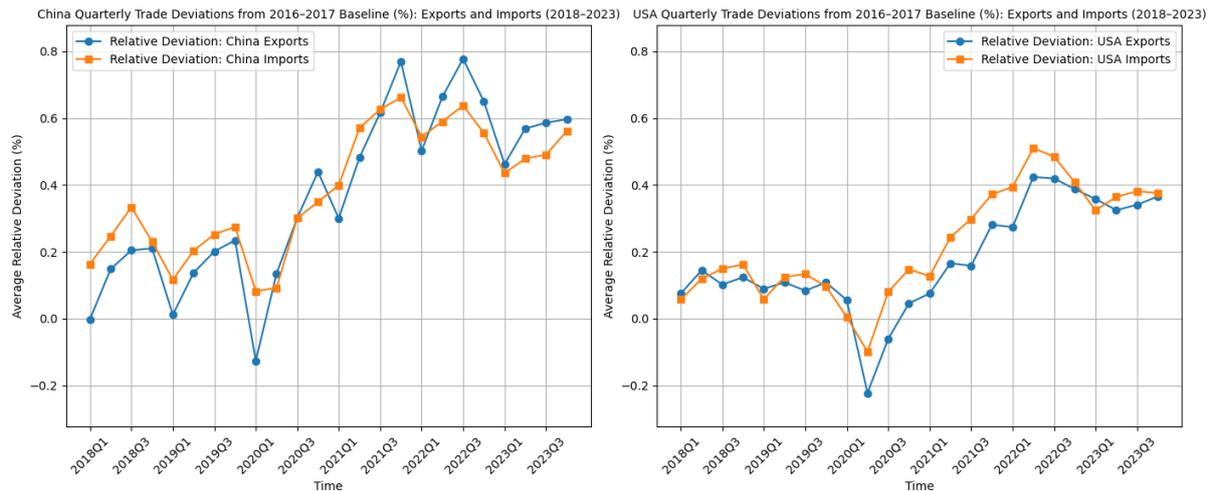

**Figure 1.** Quarterly deviations in China's and the United States' import and export trade value relative to 2016–2017 baselines. Trade turnover from 2018 to 2023 is expressed as average quarterly percentage deviations from the pre-trade war benchmark period.

Figure 1 illustrates the relative monthly deviations in exports and imports for China and the United States between 2018 and 2023, benchmarked against 2016–2017 averages. Both countries exhibited an overall upward trend in trade turnover during this period, yet the timing, intensity, and persistence of shifts were distinctly shaped by the sequence of global shocks: the U.S.–China trade tensions, the COVID-19 pandemic, and the Russia–Ukraine conflict.

For China, the imposition of U.S. tariffs in 2018 did not result in a lasting contraction of export activity. On the contrary, export volumes maintained a cyclical but upward trajectory, demonstrating notable resilience to trade frictions. A temporary collapse occurred at the onset of the COVID-19 pandemic in early 2020, with exports sharply declining until April. However, a swift recovery followed, with trade rebounding to pre-pandemic levels by mid-2020 and accelerating further in 2021. The final quarter of 2021 marked the peak of China's export expansion. Although growth rates moderated in 2022, they remained above pre-crisis levels. In 2023, export activity experienced a modest decline in the first quarter, followed by a mid-year recovery and a period of relative stabilization through the end of the year. China's imports followed a broadly similar pattern to exports during 2018–2019, outpacing export growth during that phase. However, the pandemic's disruptive effects on imports were more delayed and peaked in May 2020. From June 2020 onward, import levels resumed a steady climb, though export growth increasingly outstripped that of imports, particularly in the post-pandemic recovery phase.

By contrast, the United States exhibited a different pattern of trade volatility. From 2018 to 2019, export growth remained modest, with relative deviations fluctuating between 0 and 0.2 percentage points. The onset of the COVID-19 pandemic precipitated a sharp and sustained decline in exports, with the most pronounced contraction occurring between April and September 2020. A recovery began in early 2021, and by mid-2021, the growth rate of U.S. exports had exceeded pre-pandemic levels. Following the escalation of geopolitical tensions in 2022, trade dynamics again became unstable, initially marked by a notable slowdown in export growth, which was subsequently followed by a gradual rebound throughout 2023. Throughout the six-year period, U.S. import growth consistently outpaced that of exports. Import activity rebounded swiftly after the pandemic, achieving full recovery by the third quarter of 2020. In the period following the Russia–Ukraine conflict, import levels rose further, reaching a peak in the second quarter of 2022. However, this upward momentum did not persist, as import growth decelerated sharply in 2023.

# Trade Event Study Analysis: Spatio-temporal reallocations over global shocks

**(A)**

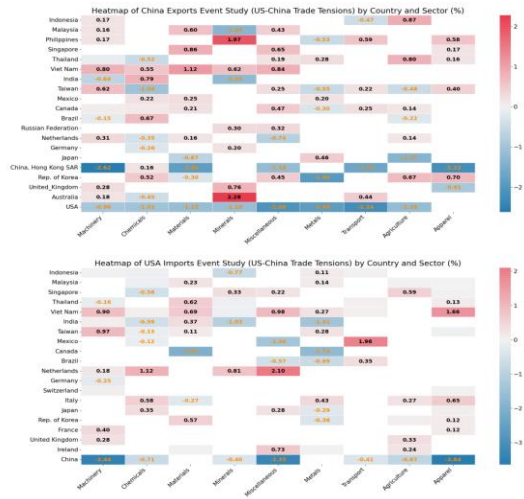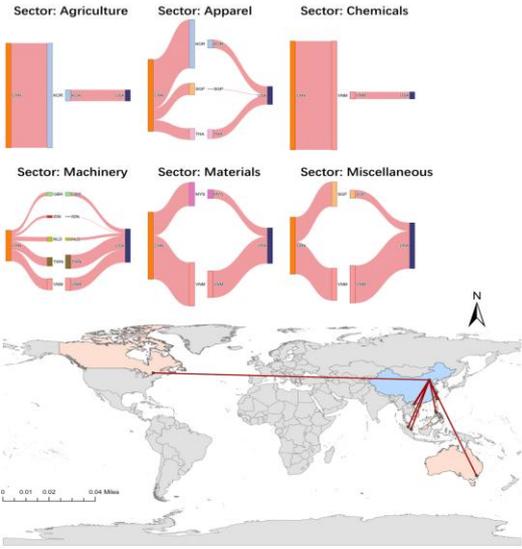

**(B)**

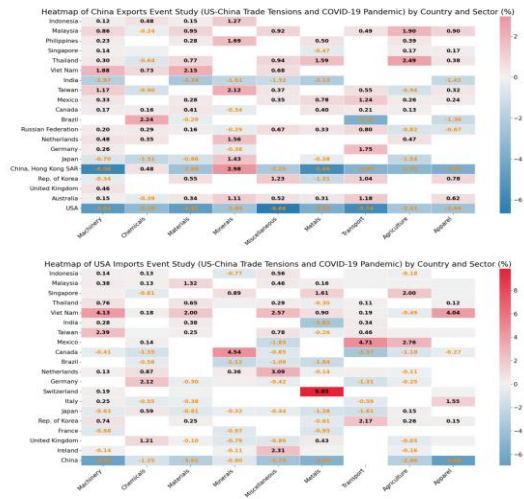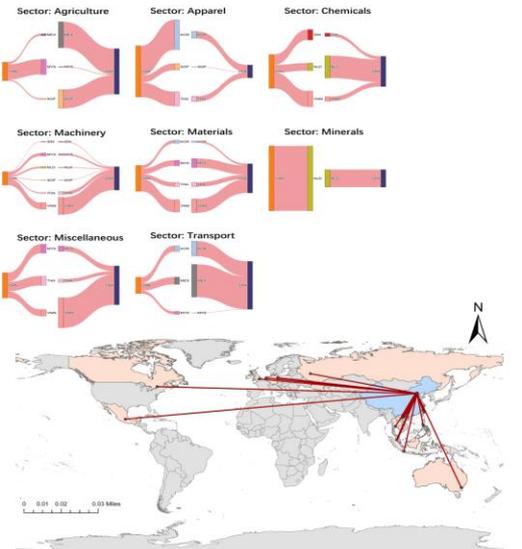

**(C)**

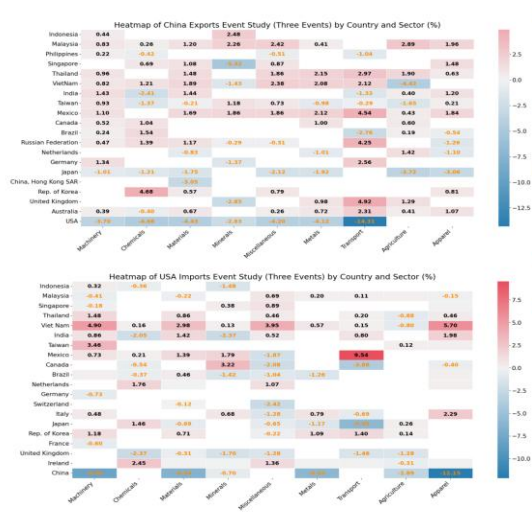

**Figure 2. Event-study estimates of sectoral shifts in China's exports and U.S. imports under successive global disruptions.** Panels (A)–(C) report changes in monthly average export market shares from mainland China and corresponding import shares to the United States across three major global disruptions: (A) the onset of the U.S.–China trade tensions, (B) the compounded impact of the trade war and the COVID-19 pandemic, and (C) the cumulative influence of the trade war, the pandemic, and the Russia–Ukraine conflict. In each panel, the upper heatmap presents event-study estimates for Chinese exports, while the lower heatmap shows matched results for U.S. imports by country and sector. The top-right diagrams identify sectoral combinations that experienced statistically significant positive shifts in both Chinese exports and U.S. imports, highlighting jointly impacted nodes of bilateral trade strengthening. The lower-right maps visualize aggregated changes in Chinese exports across five key sectors—chemicals, machinery, materials, miscellaneous goods, and transport—to show the evolving spatial structure of positive trade responses. Estimates are derived from event-study regressions that compare each treated country–sector pair against a synthetic control group composed of all other destinations. The pre-event baseline is defined as January 2016 to September 2018. Post-event windows are October 2018 to December 2019 (A), January 2020 to January 2022 (B), and February 2022 to December 2023 (C). Coefficients represent statistically significant deviations from the pre-event baseline ($p < 0.10$). Estimated effects of control variables are reported in Supplementary Tables S1–S12.

**Panel (A)** illustrates the shifts in mainland China's exports and U.S. imports following the onset of the U.S.–China trade tensions. On the export side, China's trade with the United States contracted across nearly all sectors, with the steepest reductions recorded in metals, miscellaneous goods, and transport equipment. A comparable decline is observed in exports to Hong Kong SAR, where agriculture and apparel experienced the most significant losses. In contrast, exports to several Southeast and East Asian partners showed differentiated responses: machinery, materials, and miscellaneous goods registered moderate gains, particularly in Viet Nam, Thailand, and Malaysia. These positive adjustments extended to exports to Mexico, Canada, and Australia, suggesting a partial reorientation of demand toward alternative markets. U.S. imports from China declined sharply and consistently across most product categories, especially in apparel, machinery, and miscellaneous goods—sectors closely linked to labor-intensive and intermediate manufacturing. These changes reflect early signs of supply chain realignment and decoupling. Imports from other American partners, including Brazil, Canada, and Mexico, also contracted in key sectors such as materials and minerals. In contrast, increased import activity was concentrated in Asia and parts of Europe. Notable gains were observed in apparel and machinery from Thailand, Korea, and Viet Nam, with the latter emerging as a prominent alternative sourcing hub. Rising imports of materials from Malaysia, Taiwan, and Thailand point to growing regional diversification efforts. Modest increases in European imports—particularly in chemicals, metals, and miscellaneous goods from Italy and the Netherlands—indicate early transatlantic adjustments in supplier portfolios. The top-right diagram highlights the country–sector combinations that recorded simultaneous positive effects in both China's exports and U.S. imports. Viet Nam notably exhibited gains across four sectors—chemicals, machinery, materials, and miscellaneous goods—underscoring its central role in the shifting trade landscape. Most dual-positive cases were concentrated in Asian economies. When aggregating the five core sectors (chemicals, machinery, materials, miscellaneous, and transport equipment), China's export increases were geographically concentrated in countries such as Thailand, Viet Nam, Singapore, and the Philippines, alongside partners including Canada and Australia.

**Panel (B)** illustrates trade shifts under the compounded effects of the U.S.–China trade tensions and the COVID-19 pandemic. Compared to Panel (A), the contraction in Chinese exports to the United States and Hong Kong SAR intensified across multiple sectors, including machinery, materials, miscellaneous goods, metals, and transport equipment. In contrast, exports to several Asian economies—notably Indonesia, Malaysia, the Philippines, Thailand, Singapore, and Viet Nam—expanded broadly, with machinery registering the most pronounced growth. These patterns suggest a regional reconfiguration of demand in response to global supply disruptions, along with China's increasingly central role in intra-Asian trade. Machinery exports to Canada, Mexico, and select European destinations also displayed resilience or gains, reflecting a partial redirection of trade flows. On the U.S. import side, positive deviations became more widespread, with increases largely concentrated in Asia, Mexico, and Australia. Machinery imports rose substantially from India, Indonesia, Malaysia, Singapore, Thailand, Taiwan, Korea, and Viet Nam—the latter again recording the broadest gains across sectors. At the same time, import declines from China deepened, particularly in apparel (−6.96%), machinery (−6.66%), and minerals (−5.20%). Imports from Canada also weakened across agriculture, chemicals, and machinery, although minerals remained stable, possibly due to strategic sourcing adjustments for critical raw materials. In Europe, negative shifts became more visible, especially in materials. One exception was Switzerland, where metal imports surged (+9.85%), suggesting selective shifts in high-value manufacturing inputs. The upper-right diagram highlights a growing number of country–sector combinations exhibiting simultaneous gains in both U.S. imports and Chinese exports following the pandemic. Mexico showed dual positive effects in agriculture and transport equipment. Malaysia experienced gains across a wider range of sectors, including agriculture, machinery, materials, miscellaneous goods, and transport. Korea registered dual gains in apparel, materials, and transport, while Thailand and Viet Nam maintained strong co-positive signals in several industrial sectors, notably apparel, chemicals, machinery, materials, and miscellaneous goods. In Europe, the Netherlands began to emerge with positive shifts in chemicals and minerals. When aggregating results for five key sectors (chemicals, machinery, materials, miscellaneous goods, and transport equipment), the spatial footprint of China's export gains expanded markedly. Positive effects became more geographically dispersed, extending across additional Asian economies such as Malaysia, the Philippines, and Indonesia, as well as into Europe—including the Netherlands and Russia— and to American partners such as Mexico.

**Panel (C)** presents trade reconfigurations under the compounded effects of the U.S.–China trade tensions, the COVID-19 pandemic, and the Russia–Ukraine conflict. Relative to previous periods, China's exports demonstrated broader and more consistent positive shifts across multiple sectors—particularly machinery, chemicals, materials, and miscellaneous goods. Aside from persistent contractions in trade with the United States and Japan, Chinese machinery exports registered widespread gains across nearly all major trading partners. Southeast Asian economies—especially Malaysia, Thailand, and Viet Nam—exhibited continued and expanded growth across a diverse range of sectors, underscoring China's strengthening role as a regional hub for capital and intermediate goods. India, which previously showed negative responses during the pandemic, transitioned to significant export gains from China in machinery, materials, agriculture, and apparel under the tripartite shock. Notably, China's transport equipment exports to Europe rebounded, hinting at emergent substitution strategies amid escalating geopolitical and logistical uncertainties. In contrast, exports to the United States remained persistently suppressed, with transport equipment experiencing the steepest and most prolonged declines—suggestive of structural disengagement in technologically and logistically complex supply chains. U.S. import patterns continued to reflect intensified decoupling from China. Imports from mainland China declined further, with apparel (−12.15%), machinery (−8.06%), and metals (−6.03%) registering substantial reductions. At the same time, Mexico emerged as a principal beneficiary of reshoring dynamics, with marked increases in transport equipment (+9.54%) signaling a broader North American realignment. Within Asia, substitution pathways remained strong, as imports of machinery, materials, and miscellaneous goods rose from key regional suppliers. Viet Nam retained its position as a central alternative, posting significant gains in nearly all categories except agriculture. Meanwhile, the United Kingdom faced sustained import declines, whereas Italy maintained upward momentum across several sectors, reflecting heterogeneous supplier repositioning within Europe. The upper-right diagram highlights country–sector combinations that exhibited dual positive effects—simultaneous increases in both U.S. imports and Chinese exports—under the full set of disruptions. ASEAN members, including Malaysia, Thailand, Viet Nam, Indonesia, and Singapore, remained prominent, with Viet Nam showing co-positive shifts in six sectors (chemicals, machinery, materials, metals, miscellaneous goods, and transport equipment), and Thailand in five. Additional gains emerged from other Asian economies such as India, Taiwan, and the Republic of Korea, each registering dual-sector gains. Compared with Panel (B), Mexico and Brazil played a more prominent role in this final stage, reflecting intensified diversification strategies. When aggregating across the five major industrial sectors, the spatial footprint of China's export strength became even more concentrated in key markets—most notably Mexico, Viet Nam, Thailand, and Malaysia.

## GVC Participation and Position: Deepening and embedded triangular trading relationships

### Geoeconomic divergence in GVC participation across major economies and regional blocs

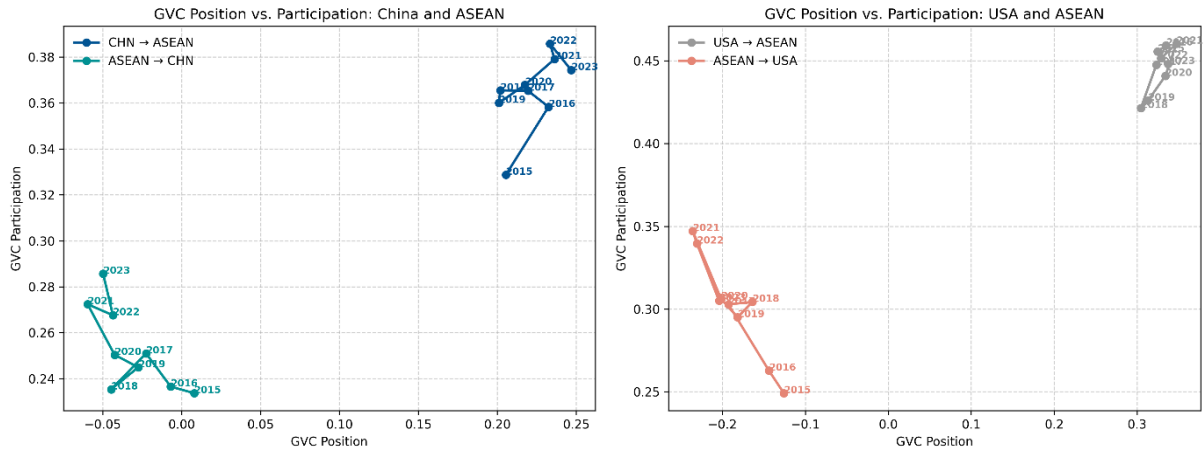

**Figure 3. GVC participation and position in exports between China, the United States, and ASEAN economies (2015–2023).** Each panel traces the annual trajectories of GVC participation and position for four directional trade flows: (a) China → ASEAN, (b) ASEA → China, (c) United States → ASEAN, and (d) ASEAN → United States. GVC participation is measured as the share of gross exports involved in cross-border production, combining both forward and backward linkages. GVC position reflects relative upstreamness, proxied by the ratio of domestic value added embedded in partners' exports to foreign value added used in own production. Trends are derived from a decomposition of gross exports into value-added components and highlight the asymmetric restructuring of regional supply chains in response to global trade shocks.

Figure 3 traces the evolution of GVC participation and position across four regional trade corridors involving China, the United States, and ASEAN member economies from 2015 to 2023. GVC participation is measured as the share of gross exports involved in either upstream or downstream cross-border production, combining both forward and backward linkages. The position index captures the relative upstreamness of a country's exports, defined by the ratio of its domestic intermediates used in others' exports to its reliance on imported intermediates in its own production.

China's exports to ASEAN demonstrate a steady upward trajectory in GVC participation over the observed period, interrupted only by a temporary contraction in 2019, coinciding with the COVID-19-induced global supply chain shock. This trajectory reflects an expanding role for Chinese intermediate goods within regional re-export production processes. Concurrently, China's GVC position in this corridor shifted progressively upstream, suggesting intensified embedding of Chinese inputs in earlier stages of multi-country value chains and reinforcing China's structural centrality within Southeast Asian manufacturing networks.

In contrast, U.S. exports to ASEAN exhibit relatively stable GVC position and participation levels over the period 2015–2023. A temporary decline in GVC participation is observed during 2018–2019, coinciding with the onset of trade tensions and suggesting a shift in U.S. exports toward direct final consumption or localized production within ASEAN economies, rather than continued embedding in cross-border value chains. However, both indicators recovered to pre-disruption levels from 2020 onwards, indicating a partial restoration of the United States' functional role in ASEAN's production networks.

Exports from ASEAN to China reveal a more stable downstream position but are accompanied by a gradual increase in GVC participation, particularly from 2017 forward. This asymmetry indicates that although the functional role of these exports within value chains remained relatively unchanged, a rising share is being integrated into Chinese re-exports—highlighting the consolidation of China as a re-export hub within the region without a corresponding upgrade in supplier economies' functional positions.

A pronounced structural shift is also evident in the export flow from ASEAN to the United States. Between 2015 and 2021, this corridor experienced a sustained rise in GVC participation alongside a marked shift toward a more downstream position consistent with an export structure increasingly dominated by final-stage assembly activities. These dynamics suggest a growing dependence on ASEAN economies for low-value-added manufacturing of consumer goods destined for the U.S. market. Notably, this pattern reversed in 2023, with GVC participation declining sharply while position remained persistently downstream. This decoupling may signal abrupt shifts in U.S. sourcing strategies—potentially triggered by geopolitical tensions, input price volatility, or a pivot toward bilateral and nearshoring arrangements in the aftermath of the Russia–Ukraine conflict.

**Sector-Economy level: Industry-specific realignment in value-added trade networks**

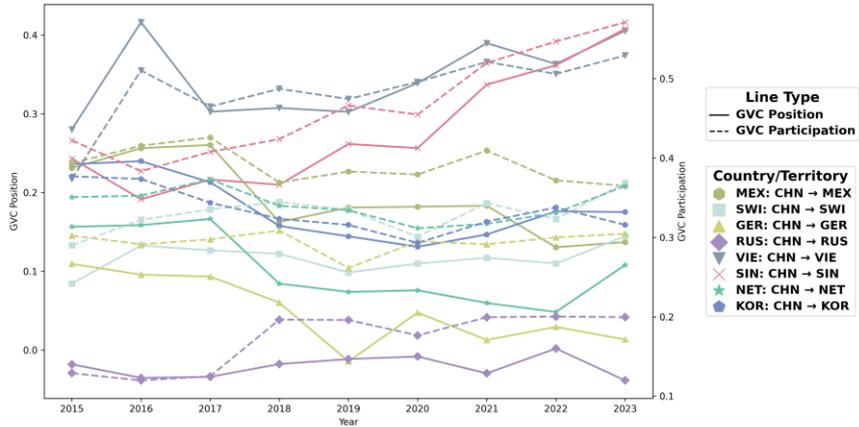
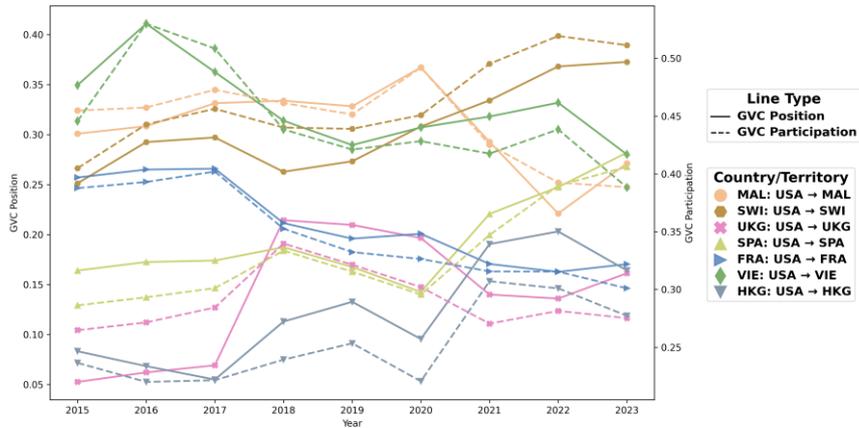
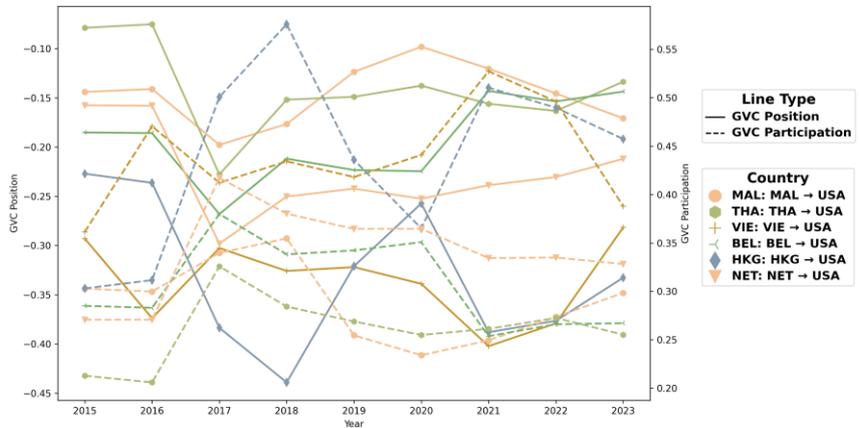
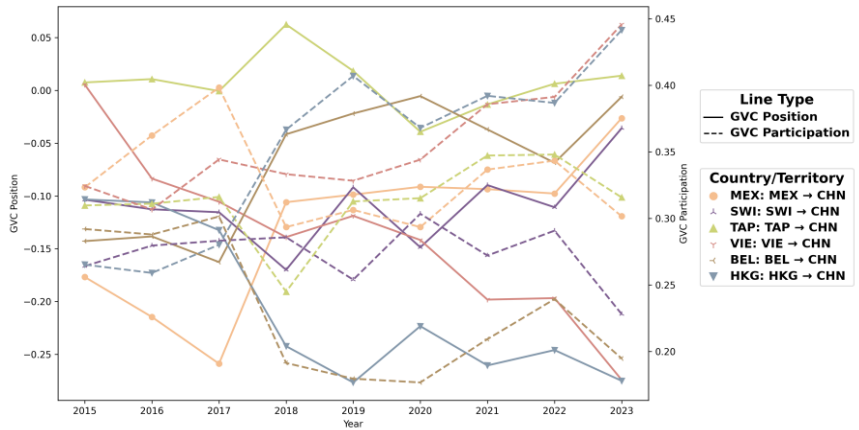

**Figure 4. Shifting patterns of GVC participation and position across selected manufacturing sectors and countries (2015–2023).** This figure depicts temporal changes in GVC participation and position for selected partner economies across major manufacturing sectors during 2015–2023. Results are shown for aggregate manufacturing (ADB MRIO sectors C3–C16); sector-specific trajectories for C14 (electrical and optical equipment) and C15 (transport equipment) are provided in the Supplementary Appendix.

Figure 4 traces the evolution of GVC participation and position from 2015 to 2023 across selected manufacturing sectors and key bilateral trade flows. For aggregate manufacturing (C03–C16), a broad upward trend is observed in China's GVC participation in its exports to Singapore, Viet Nam, Switzerland and Russia—interrupted only by the transient contraction in 2020 during the global COVID-19 lockdown. This pattern reflects a higher degree of GVC embeddedness in Chinese manufacturing exports to these destinations, driven by both increased use of foreign inputs in domestic production and greater use of Chinese intermediate goods in partner countries' exports. In parallel, modest upward movements in GVC position are detected in exports to Singapore and Viet Nam, suggesting gradual upstream shifts and enhanced functional embedding in regional value chains. These dynamics are consistent with sectoral patterns documented in the event study analysis (Fig. 2), particularly for machinery and transport equipment.

These economies exhibit divergent trajectories in their export integration with the U.S. value chain. Over the broader period, Malaysia, Viet Nam, and Hong Kong SAR display steadily rising GVC participation alongside declining upstream positioning—indicative of deepening specialization in downstream segments such as component integration and final assembly. However, between 2018 and 2020, participation levels dipped notably, reflecting the impact of U.S.–China trade tensions and pandemic-induced supply shocks, as well as potential reshoring responses within the U.S. market. For Viet Nam and Thailand, the pattern shifts markedly after 2022: participation drops sharply while position indices rise, suggesting a partial upstream relocation. This shift may be driven by supply chain realignment or demand-side adjustments under prolonged geopolitical and trade uncertainties.

In sector C14 (electrical and optical equipment, Figure S3), China's exports to both Singapore and Viet Nam remain consistently upstream, with GVC participation levels higher than those of their U.S. counterparts. Viet Nam's C14 exports to the United States rose sharply in participation between 2015 and 2021, accompanied by a decline in position—signaling intensified downstream specialization. However, this trend attenuates after 2021, as exports become increasingly oriented toward U.S. final demand rather than intermediate processing. In contrast, Viet Nam's exports of C14 goods to China show sustained increases in both participation and downstream positioning, consistent with continued integration into China-centric production systems.

For transport equipment (C15, Figure S4), China's exports to Thailand show limited movement prior to 2020, followed by a sharp increase in both GVC participation and position after 2021—indicative of functional upgrading and a shift toward more upstream roles. Thailand's own C15 exports to both China and the U.S. initially follow a comparable trajectory: declining GVC position coupled with increasing participation, consistent with a move toward final assembly roles. The patterns, however, diverge during post-2021: GVC participation in exports to the U.S. drops abruptly, while the decline in exports to China is more gradual. Meanwhile, China's C15 exports to Viet Nam show an anomalous participation spike in 2016, but excluding this outlier, exhibit a clear upward trajectory in both participation and position. Viet Nam's C15 exports to China declined significantly in participation between 2016 and 2019 but then recovered steadily after 2020. In contrast, exports to the U.S. show a consistent rise in GVC participation and a gradual downstream shift, reflecting Viet Nam's increasing concentration in final-stage transport equipment production. From 2022 onward, however, participation levels begin to decline, potentially reflecting shifting demand dynamics and logistics-related adjustments in the wake of geopolitical disruptions.

# Discussion

This study offers a U.S.–China–centered perspective on the global reconfiguration of bilateral trade and GVC linkages from 2016 to 2023, examining how major external disruptions—including the U.S.–China trade tensions, the COVID-19 pandemic, and the Russia–Ukraine conflict—reshaped their trade dynamics and affected third-party trading partners. We leverage monthly bilateral trade data across the top 20 trading partner countries with the U.S., and China in nine sectors, together with multi-regional input–output tables for GVC decomposition. We reveal the evolving patterns of US and China in bilateral trade and their functional participation and positioning in GVCs across different partner economies and products, as well as describe how this reflects their supply chain strategies in the observed friendshoring, nearshoring, and reshoring.

Though we expected that U.S. and China would expand new economic sectors and markets to reduce interdependence from each other, both countries demonstrate totally distinct supply chain patterns over those destructively compounded shocks. Rather than signaling a trend towards reducing dependence on Chinese supply chains, the available data implies a consistently increasing Chinese exports worldwide to avoid the reliance on the U.S. market. Increased tariffs cause the U.S. to import less directly from China, with a corresponding rise in import share from friendshoring and nearshoring countries (i. e., ASEAN, Mexico), who meanwhile show the significant rise in imports from China. This reveals a growing triangular export relationship from China + 1 (i.e., ASEAN) to the United States.

In addition to the triangular relationship in trade volume, our GVC decomposition reveals a deepening embeddedness of Chinese supply chains in cross-border production structures, particularly through forward linkages into third-party economies. Between 2015 and 2023, China's manufacturing exports to countries such as Viet Nam, Singapore, and Thailand displayed a sustained rise in GVC participation, indicating that a growing share of Chinese value added is being embedded upstream in partner economies' re-exported goods—especially in machinery and transport equipment. As we expected, this implies that China is not merely circumventing direct export restrictions to the U.S. but is actively embedding itself into the core production stages of third-party supply chains when facing a series of external shocks.

At the same time, third-party countries display heterogeneous functional shifts within GVCs that reflect both opportunity and constraint. Economies such as Viet Nam and India initially expanded their downstream assembly roles—likely benefiting from U.S. trade diversion policies—but their recent movements toward more upstream positioning and lower GVC participation may reflect constraints in absorbing economic complexity[43], rising costs, or broader geopolitical uncertainties.

In contrast, the United States has seen a progressive weakening of its role within regional value-added circulation, particularly in its exports to ASEAN economies. Declines in both participation and position indicate a retreat from intermediate production functions, consistent with the broader narrative of U.S. industrial repositioning and reshoring ambitions. Yet, the persistent downstream dependence on ASEAN economies, especially for final-stage consumer goods, highlights the limits of near-shoring and friend-shoring strategies when not accompanied by parallel investments in upstream capabilities.

These findings carry significant implications for trade policy and industrial strategy in an era of geopolitical fragmentation. Despite efforts to decouple and redirect supply chains through tariffs, reshoring, or friendshoring, the persistent embedding of Chinese value added in third-party exports suggests that such strategies may face structural constraints. The deeply rooted integration of China within the upstream segments of GVCs—particularly through forward linkages into Southeast Asia—reflects not only cost arbitrage, but also technological scale, logistics superiority, and institutional efficiency. According to the GVC governance framework, these structural capabilities position China as a lead firm or critical platform node in modular and relational value chains. As such, attempts at trade reconfiguration without concurrent development of comparable capacities in alternative hubs are unlikely to yield genuine realignment.

For emerging economies, the asymmetric functional shifts observed in our analysis highlight both the potential and the fragility of their GVC participation. Many third-party countries initially experienced increased forward integration—serving as final assembly hubs within the China–U.S. triangle—but the recent flattening or reversal of GVC indicators reveals a vulnerability to both exogenous shocks and endogenous constraints. Literature on GVC upgrading suggests that sustaining functional mobility requires not just trade diversification, but also capability building in design, process innovation, and standards compliance[44,45]. In the absence of these conditions, these countries risk falling into a "functional lock-in," where they remain trapped in low-margin, downstream activities that are easy to relocate but hard to upgrade[46].

The results also prompt a rethink of supply chain resilience. Traditional interpretations equate resilience with geographic diversification or sourcing flexibility[47]. However, our findings suggest that structural embeddedness—i.e., occupying critical upstream positions and building forward linkages into re-export platforms—may offer more sustainable forms of control over value chains. As Christopher and Peck [48] argued, resilience depends not merely on spatial dispersion, but on visibility, agility, and collaboration across tiers. In this context, we argue that supply chain resilience is shaped by the interaction of three key dimensions: the level of GVC participation, which determines exposure volume; the functional position within the value chain, which influences vulnerability and substitutability; and the country's capacity to re-couple in the post-shock landscape, which depends on market diversification, economic complexity, and institutional capability (i.e., upstream investment). For example, China's ability to reposition itself within third-party production systems points to a reconfiguration of resilience: from redundancy to relatedness[49]. Conversely, the United States' erosion in both participation and position underscores the challenges of restoring upstream capabilities without sustained industrial reinvestment.

Despite the richness of multi-source trade and input–output data, this study is subject to several limitations. First, while our GVC decomposition captures the structural embedding of economies within value-added trade flows, the analysis remains confined to sector-aggregated MRIO data, which may mask firm-level heterogeneity and intra-industry dynamics critical to understanding strategic supply chain decisions. Second, the interpretation of upstream and downstream shifts relies on proxies that are abstract from firm strategies, technological intensity, and institutional arrangements, which may influence the actual functional roles of countries. Third, the triangular trade relationships observed—such as China's indirect exports to the U.S. via ASEAN—are inferred from macro trade linkages rather than firm-level transaction networks, which limits causal attribution. Lastly, although our event-based analytical framework provides temporal clarity, the overlapping nature of trade war, pandemic, and geopolitical conflicts may blur the discrete impact of each disruption. These caveats suggest the value of future research integrating micro-level supply chain data, firm-level surveys, and higher-resolution sectoral classifications.

# Methods

*Data*

This study integrates multiple data sources to investigate the evolving structure of international trade and GVCs, with a focus on bilateral and third-party trade relations between the United States, China, and their key partners from 2016 to 2023.

Bilateral trade data were primarily drawn from the United Nations Comtrade database, which compiles standardized merchandise trade statistics from nearly 200 economies, covering over 99% of global trade. For the United States, we used mirror data based on U.S.-reported exports and imports by partner country; for China, we relied on official Chinese-reported trade statistics to ensure consistency and reporting accuracy. Trade flows are recorded at the 2-digit level of the Harmonized System (HS) classification, allowing for cross-sectoral analysis while maintaining manageable aggregation. In line with recent literature [50], we categorize the dataset into nine sectors: agriculture, apparel, chemicals, materials, machinery, metals, minerals, transport equipment, and miscellaneous manufacturing. The final dataset includes over 1.7 million trade records for the U.S. and 1.8 million for China, with monthly observations capturing trade values, HS codes, directionality, and partner country.

To contextualize trade dynamics within broader structural and geopolitical shifts, we integrated several macro-level indicators. These include:

- **Lagged population and GDP per capita growth** [51,52] to proxy for demand-side pressures and development trajectories;
- **Geopolitical distance**, measured via ideal point distances in UN General Assembly voting records [53], as a time-varying proxy for bilateral political alignment;
- **Socioeconomic condition and investment profile,** captured using the International Country Risk Guide (ICRG) composite indices [54], which quantify country-specific institutional and economic vulnerabilities on a 0–4 scale.

To analyze trade in value-added terms, we employed the Asian Development Bank's Multi-Regional Input–Output (ADB-MRIO) database. This dataset comprises annual input–output tables for 62 economies and one aggregated Rest-of-World region, covering 35 industrial sectors. It spans the full 2015–2023 period and is among the most comprehensive and up-to-date MRIO databases available for Asia-Pacific and global supply chain analysis. The ADB-MRIO data underpin the application of the Wang–Wei–Zhu (WWZ) decomposition framework[55], a widely adopted methodology for tracing value-added flows in GVCs. The WWZ approach decomposes gross exports into 16 value-added components, enabling the distinction between forward and backward GVC linkages, as well as identifying countries' functional roles in cross-border production chains. By combining monthly trade records with structural input–output linkages, this study offers a multilayered perspective on international trade reconfigurations, allowing us to trace the redistribution of production stages and value-added contributions in response to geopolitical and pandemic-related disruptions.

**Methods**

**Event-Study Analysis**

To examine the causal impact of major geopolitical shocks on bilateral trade structures, we employ a multi-scale event study framework grounded in panel data econometrics. This approach enables us to trace the temporal effects of three successive disruptions—the U.S.–China trade tension, the COVID-19 pandemic, and the Russia–Ukraine conflict—across country-sector trade flows over the period 2016–2023.

The identification strategy is anchored in a quasi-experimental design that contrasts outcome trajectories between treatment and control groups before and after each event. Specifically, we estimate a two-way fixed-effects panel model that captures both country-sector heterogeneity and time-specific shocks. The treatment status is assigned based on ex-ante exposure to each disruption, with carefully matched counterfactuals used to construct a credible comparison baseline. This allows for the detection of dynamic response patterns across partner countries and product categories while mitigating concerns of unobserved confounders.

We employed innovative methodology to analyze the value chain reallocation by combining the power of iteratively multi-scale event studies and visualization. To conduct our event study analysis, we first utilize a difference-in-differences OLS regression with 'country'/'region' fixed effects. This approach mitigates the omitted variable issues associated with unobservable heterogeneity and time-specific factors. The resulting regression coefficients were assessed at 10% or lower significance levels to determine the most appropriate one for different data pairs. Compared to other empirical works, such as the gravity model and those built directly from specific models, our approach is more robust and less sensitive to variable omissions, data outliers, and theoretical oversimplifications. Our iterative approach allowed for a comprehensive examination of economic sectors at both continental and national or regional levels, contributing to a deeper understanding of the nuanced dynamics and impacts of the events analyzed. Then, to present our massive results comprehensively, we rely on various visualization representations to illustrate the relative changes over numerous metrics.

By accounting for the trade value variable and incorporating geographically fixed effects, our study seeks to provide credible estimations of the numerical consequences of these events on U.S. and China's export/import trade. The model equation is as follows:

$$Trade_{(i,t)} = \alpha * Treat_i + \beta * Post + \gamma * (Treat_i * Post) + \theta_{(i)} + \lambda_1 * PopGrowth_{i,t-1} + \lambda_2 * GDPGrowth_{i,t-1} + \lambda_3 * GeoDist_{i,t} + \lambda_4 * SocioCond_{i,t} + \lambda_4 * InvestPofile_{i,t} + \varepsilon_{(i,t)}$$

where the U.S. and Chinese exports/imports trade share value ($Trade_{(i,t)}$) is the explained variable. $Treat_i$ is a dummy variable that denotes a focal (treatment) region or country (set to 1) from the control regions or countries (set to 0). $Post$ is set to 1 for the event period and 0 for the comparison or benchmark period. $Treat_i \times Post$ is a core explanatory variable, the interaction variable of $Treat_i$ and $Post$.

Country/region fixed effects ($\theta_{(i)}$) are included to control for time-invariant characteristics specific to each country or region. The model also incorporates a set of control variables to account for potential confounding factors, including lagged population growth ($PopGrowth_{i,t-1}$), lagged per capita GDP growth ($GDPGrowth_{i,t-1}$), geopolitical distance ($GeoDist_{i,t}$), socioeconomic conditions ($SocioCond_{i,t}$), and investment profile ($InvestPofile_{i,t}$). These control variables ensure that the estimated effect of the treatment is not biased by demographic, economic, geopolitical, or risk-related influences.

Calculated by the model, the coefficient of this variable, that is $\gamma$, measures the effects of the events on U.S. and China's imports/exports trade. Without fixed effect variables, $\gamma$ measures the differences between the change in the focal country's or region's trade share and that of benchmarking countries or regions after or before an event [56]. Region/country -fixed effects tend to subsume the treatment/post-dummy variables and deviate the $\gamma$ slightly away from the exact value from the difference-in-differences. Despite the deviation, social scientists still consider that the interaction coefficient from this regression can reasonably approximate the difference-in-differences in the sample.

**Value Decomposition Framework and Indicator System**

Global trade in intermediate and final goods involves complex value-added linkages across economies, requiring a structured decomposition framework to disentangle the origin, transformation, and redistribution of value within supply chains. This study applies ADB-MRIO to systematically decompose the bilateral trade of 63 economies from 2016 to 2023. The decomposition follows the WWZ framework 42,44, segmenting total exports into 16 distinct components (T1–T16) that capture the creation, absorption, and reallocation of value-added across production networks.

Unlike conventional studies that focus on absolute trade values, this research employs a ratio-based approach, expressing each value-added component as a proportion of total exports to facilitate cross-country comparisons and structural analysis over time. This methodological refinement enhances the interpretability of trade dynamics, particularly in assessing value capture efficiency, global supply chain dependencies, and trade-induced distortions.

The total export value of metal from economy s to economy r, denoted as $E^r$, can be expressed as:

$$E^r = A^s * X^r + Y^r = T1 + T2 + \cdots + T16$$

Where $A^s X^r$ represents intermediate trade, and $Y^r$ denotes final demand exports. Each of the 16 components reflects different channels of value generation, transfer, and redistribution, forming the basis for a refined indicator system.
- T1: Domestic value added of final exports.
- T2: Domestic value added of intermediate exports absorbed by direct importers.
- T3+T4+T5: Domestic value added of intermediate exports processed by direct importers and re-exported to third countries.
- T6+T7+T8: Domestic value added returned to the exporting country via re-imports.
- T9+T10: Double Counted Domestic Value added used to produce final/intermediate exports.
- T11+T12: Direct importer's value added embedded in the source country's exports (both final and intermediate goods)
- T13+T14: Third-country value added implicitly in domestic exports.
- T15+T16: Foreign value-added double-counting terms.

To operationalize the structural characteristics of GVC engagement, two key indicators are constructed based on the above decomposition: **GVC Participation** and **GVC Position**[57]. These indicators move beyond aggregate trade statistics and capture the functional role an economy plays in international production networks.

**GVC participation** captures the intensity of an economy's involvement in international production fragmentation. Specifically, the GVC participation from a source country *s* to a partner (or reporting) country *r*, denoted as $GVC_{participation_{s \to r}}$, is defined as sum of forward participation—measured by the domestic value added in intermediate exports that are further re-exported by partner countries (T3–T8)—and backward participation—measured by the foreign and third-country value added embedded in the exporting country's gross exports (T11–T14), all expressed as a share of total gross exports:

$$GVC_{Participation_{s \to r}} = \frac{\sum_{i=3}^{8} T_i + \sum_{i=11}^{14} T_i}{\sum_{i=1}^{16} T_i}$$

This indicator reflects the dual channels through which a country participates in the GVC: as a supplier of intermediate goods to third countries (forward participation) and as a user of imported intermediate goods for its own exports (backward participation).

GVC Position reflects the relative upstreamness of a country's exports in global production networks. The GVC position from source country *s* to reporting country *r*, denoted as $GVC_{position_{s \to r}}$, is conducted following Koopman et al.[58] as the difference between the logarithm of forward participation and the logarithm of backward participation, providing a normalized measure of functional positioning:

$$GVC_{Position_{s \to r}} = \ln\left(1 + \frac{\sum_{i=3}^{8} T_i}{\sum_{i=1}^{16} T_i}\right) - \ln\left(1 + \frac{\sum_{i=11}^{14} T_i}{\sum_{i=1}^{16} T_i}\right)$$

A higher value of GVC Position indicates a more upstream role—typically associated with the supply of core intermediate goods—while a lower value reflects a downstream role centered on assembly or final-stage processing. Together, these indicators enable a nuanced understanding of countries' evolving trade functions, structural integration, and value-added capture within global production networks.